# Single-Beam Coherent Raman Spectroscopy and Microscopy via Spectral Notch Shaping

Ori Katz, Jonathan M. Levitt, Eran Grinvald, and Yaron Silberberg

*Department of Physics of Complex Systems, Weizmann Institute of Science, Rehovot, 76100 Israel*

E-mail: ori.katz@weizmann.ac.il

**Raman spectroscopy is one of the key techniques in the study of vibrational modes and molecular structures. In Coherent Anti-Stokes Raman Scattering (CARS) spectroscopy, a molecular vibrational spectrum is resolved via the third-order nonlinear interaction of 'pump', 'Stokes' and 'probe' photons, typically using a complex experimental setup with multiple beams and laser sources. Although CARS has become a widespread technique for label-free chemical imaging and detection of contaminants, its multi-source, multi-beam experimental implementation is challenging. In this work we present a simple and easily implementable scheme for performing single-beam CARS spectroscopy and microscopy using a single femtosecond pulse, shaped by a tunable narrowband notch filter. As a substitute for multiple sources, the single broadband pulse simultaneously provides the pump, Stokes and probe photons, exciting a broad band of vibrational levels. High spectroscopic resolution is obtained by utilizing a tunable spectral notch, shaped with a resonant photonic crystal slab filter, as a narrowband, time-delayed probe. Using this scheme the entire vibrational spectrum can be resolved in a single-shot multiplexed measurement, circumventing the need for a multi-source configuration or a complex pulse-shaping apparatus. We demonstrate high-resolution single-beam micro-spectroscopy and vibrational imaging of various samples in the 300cm$^{-1}$-1000cm$^{-1}$ spectral range.**

The oscillation frequencies of molecular vibrations reflect the chemical structure and are widely used as a spectroscopic fingerprint for chemical detection and identification. One of the most efficient optical techniques to acquire the vibrational spectrum is Coherent Anti-Stokes Raman Scattering (CARS) spectroscopy [1]. CARS is a third-order nonlinear process in which 'pump' and 'Stokes' photons at frequencies $\omega_p$ and $\omega_s$, respectively, coherently excite molecular vibration at the frequency $\Omega_{vib}=\omega_p-\omega_s$ (Fig. 1a). The excited vibrational level is subsequently probed by interaction with 'probe' photons at frequency $\omega_{pr}$, generating blue-shifted Anti-Stokes photons at a frequency $\omega_{AS}=\omega_{pr}+\Omega_{vib}$. The vibrational spectrum is resolved by measuring the blue-shift of the scattered Anti-Stokes photons from the probe frequency. Since CARS spectroscopy is a nonlinear technique it benefits from intrinsic high resolution three-dimensional sectioning capability together with high efficiency at low average laser power. With the advent of powerful ultrafast lasers CARS spectroscopy has found use in numerous applications ranging from noninvasive biomedical imaging, to combustion analysis, and remote sensing [1-14].

CARS spectroscopy is typically performed as a multi-beam, multi-source technique [1-3], which is experimentally challenging to implement due to the strict requirement of spatial and temporal overlap of the excitation beams. This paradigm changed with the introduction of coherent control pulse-shaping techniques, which employ a single femtosecond pulse for CARS spectroscopy and microscopy [4-7]. In these schemes, a single femtosecond pulse simultaneously provides the necessary pump, Stokes and probe photons. The main difficulty with implementing CARS using femtosecond pulses is that the pulse bandwidth is much broader than the width and spacing of the vibrational lines, thus limiting the spectroscopic resolution [4,8]. However, through careful spectral-phase or polarization shaping, a single vibrational level can be selectively excited [4], or narrowly probed [5,6], yielding a spectroscopic resolution orders of magnitude better than the pulse bandwidth. Such single-pulse techniques are particularly attractive as only a single laser source is required, and the spatiotemporal overlap of the pump, Stokes, and probe photons is inherently maintained. Moreover, because of their high peak intensity, ultrashort pulses are favorable for a variety of nonlinear measurements. Femtosecond CARS is well suited for integration in multi-modal microscopy, combining modalities such as second- and third-harmonic generation, and multi-photon fluorescence, for label-free microscopy [9]. Since first demonstrated [4], single-pulse CARS schemes have been utilized for various applications, such as vibrational imaging [4,10,11], time-resolved chemical micro-analysis [12], and remote detection of hazardous materials from a standoff distance [13,14]. However, a major drawback of these techniques is the necessity of a programmable dynamic pulse shaper apparatus, typically a relatively complex and costly experimental setup which requires precise calibration [15]. In addition to the strict alignment and calibration requirements, conventional pulse shapers are limited in their refresh rate, spectral coverage and shaped pulse repetition rates. These limitation arise in liquid-crystal based pulse shapers (refresh rate and spectral coverage), acousto-optic based shapers (pulse repetition rates and spectral coverage), and deformable-mirrors technology (refresh rate).

In this work, we demonstrate a novel scheme for performing rapid single-pulse, single beam CARS spectroscopy, without the necessity of a complex pulse-shaping apparatus. We achieve this by shaping a narrowband notch in the pulse spectrum using a commercially available resonant photonic crystal slab (RPCS) [16,17]. Employing the RPCS as a simple and robust pulse shaping element, we utilize the tunable notch in the RPCS transmission spectrum as a narrowband, temporally extended and time-delayed probe to perform multiplex CARS spectroscopy with spectral resolution two orders of magnitude better than the pulse bandwidth. The simplicity and robustness of this single-beam, shaper-free scheme, together with its wide spectral coverage, rapid modulation capability, and straightforward alignment, are attractive for practical spectroscopic applications.

The RPCS shaped single-pulse CARS technique is analogous to conventional multiplex CARS spectroscopy (Fig.1b,c). In conventional multiplex CARS (Fig.1b) a band containing several vibrational levels is coherently excited by broadband pump and Stokes beams, and subsequently probed by a narrowband probe beam, producing blue-shifted spectral peaks in the CARS spectrum. In the RPCS single-pulse CARS technique, a narrowband probe is defined within the broad pulse bandwidth by shaping a narrow spectral notch (Fig.1c). The notch-shaped excitation pulse produces a CARS spectrum containing narrow spectral dips which are blue-shifted from the original notch frequency by the molecular vibrational frequencies. The vibrational spectrum can be easily resolved by measuring the blue-shift of the dips in the CARS spectrum from the shaped notch frequency (Fig. 1c). The spectroscopic resolution is therefore dictated by the notch spectral width rather than the full pulse bandwidth. Although intuitively depicted in the frequency domain, the RPCS notch-shaped single-pulse CARS scheme can be as easily understood in the time-domain (Fig.1c): The notch-shaped excitation pulse is a result of a destructive interference of the unshaped transform limited pulse and a temporally extended narrowband probe pulse at the notch wavelength. Since the RPCS filtering is a causal process (not necessarily the case with a pulse shaper [18]), the narrowband temporally extended probe is time-delayed to follow the femtosecond excitation (Fig.1c). The outcome is an impulsive femtosecond excitation of coherent molecular motion, followed by time-delayed, narrowband probing. The time delayed probing is effective for CARS because it does not probe at times t<0, i.e. before the vibrational coherence is excited [8,19]. The resulting vibrational spectral features in the CARS spectrum can be interpreted as the coherent interference of the CARS field induced by the ultrashort excitation pulse, and the CARS field induced by the narrowband time-delayed probe.

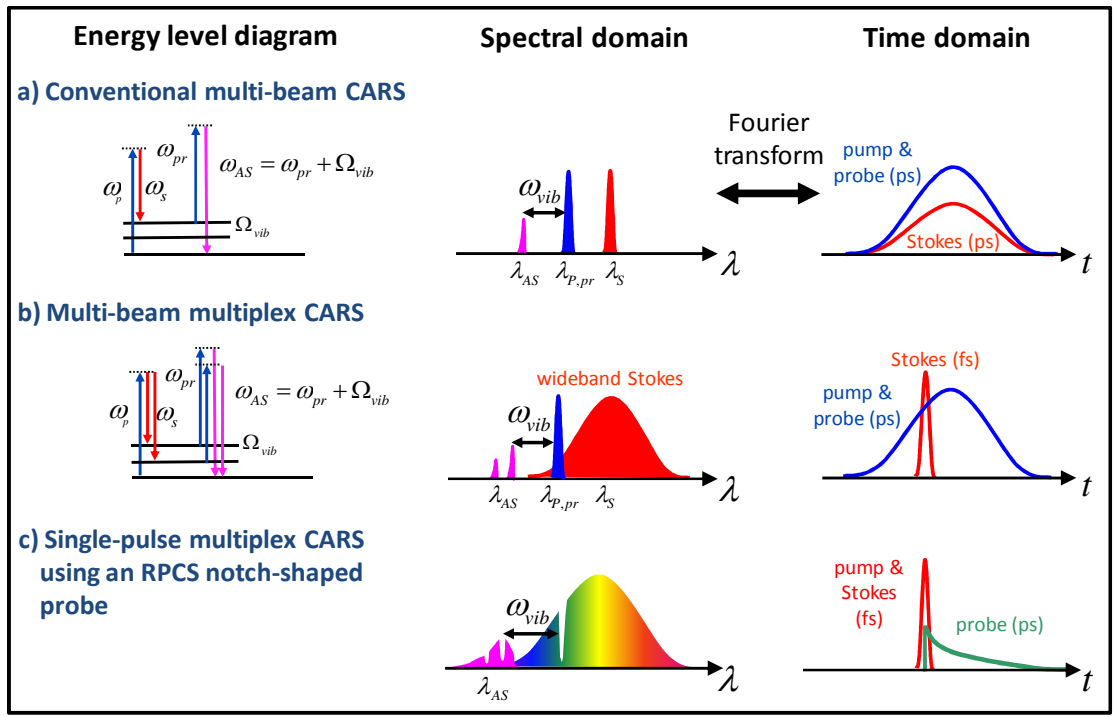

Fig.1: Various approaches for CARS spectroscopy using ultrashort pulses. The corresponding energy-level diagrams (left), and the spectral- and time-domain pictures (center and right). (a) Conventional multi-beam CARS where a single vibrational level is excited and probed by narrowband pump and a Stokes beams; (b) Multiplex CARS utilizing a wideband ultrashort Stokes pulse and a narrowband probe beam, simultaneously exciting and probing several vibrational levels; (c) Single-pulse, single beam CARS technique presented in this work, where a single femtosecond pulse is shaped with a narrow notch by a resonant photonic crystal slab filter (see Fig.2). The wideband pulse coherently excites a band of vibrational levels, and the narrowband notch serves as a time delayed temporally extended probe, yielding a spectral resolution orders of magnitude better than the pulse bandwidth. The vibrational spectrum is resolved by measuring the blue shift of the induced interference features in the CARS spectrum from the shaped notch frequency.

**Experimental setup and results**

The experimental setup for RPCS shaped single-pulse CARS technique is illustrated in Figure 2a. A wideband excitation pulse from a femtosecond oscillator (Fig. 2a-i) is spectrally notch shaped by an RPCS filter (Fig. 2a-ii). The shorter wavelengths part is blocked by a long-pass filter, as it spectrally overlap the CARS signal. The shaped pulse is focused into the sample using a microscope objective (20X, 0.4 NA, Newport), and the CARS signal is collected by a 0.5 NA condenser. The excitation signal is rejected using a sharp short-pass filter, and the CARS spectrum is measured by a fiber coupled imaging spectrometer, equipped with liquid Nitrogen cooled CCD (Jobin Yvon Triax 320).

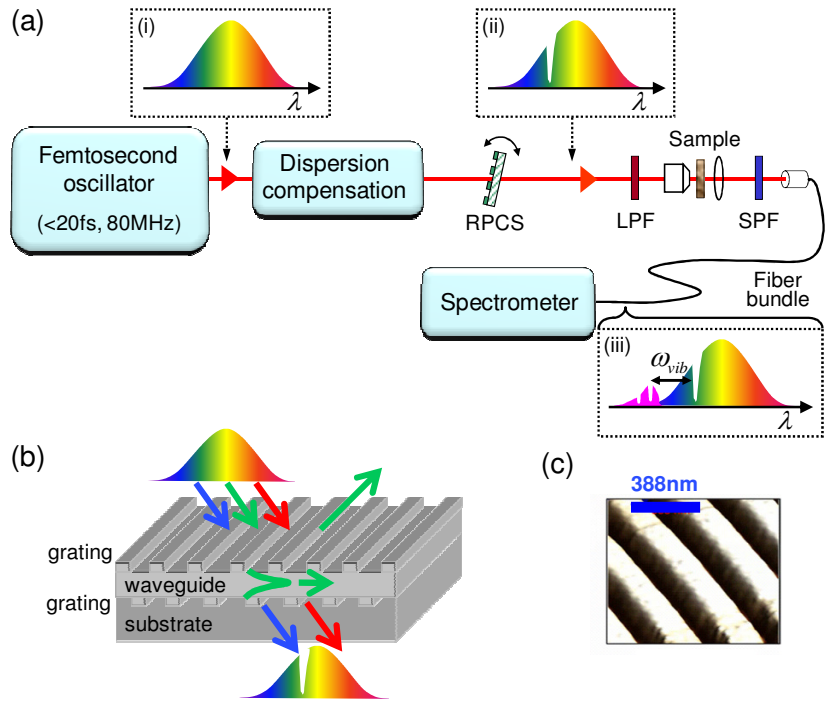

Fig. 2: Experimental setup for single-pulse shaper-less CARS using a resonant photonic crystal slab (RPCS) notch filter. (a) The optical setup: The wideband excitation pulse (i), is shaped with a tunable narrowband spectral notch by the RPCS filter (ii). The narrowband notch serves as a probe for the CARS process, generating narrow well-defined features in the CARS spectrum, which are blue-shifted from the probe by the vibrational frequencies (iii). (LPF - long-pass filter, SPF - short-pass filter); (b) A schematic diagram of the RPCS double grating waveguide structure used in this work; For a given beam incident angle, a narrow spectral band is on resonance with the RPCS, resulting in almost total reflection for the resonant wavelength and a tunable narrow (~1.3nm) dip in the transmission spectrum. (c) Atomic force microscopy measurements of the RPCS surface revealing the sub-wavelength grating with a period of 388nm and groove depth of 38nm [17].

The RPCS shaping element used in this work is a commercially available, one-dimensional, grating waveguide structure which is comprised of multiple layers (Fig. 2b) [17]: a glass substrate, a sub-wavelength grating, a thin dielectric waveguide and another sub-wavelength grating. When a light beam illuminates the RPCS at a given angle, most wavelengths are fully transmitted. However, a narrow spectral band (FWHM=1.3nm) is "on resonance" with the grating waveguide; namely, the light diffracted from the grating at these wavelengths is effectively coupled to a "guided mode" in the waveguide layer. The guided light is diffracted again by the grating and interferes destructively with the transmitted wave, resulting in full reflection of the incident wave. As a result, practically no light within the resonant band is transmitted through the RPCS, thus creating a notch in the transmission spectrum. The notch spectral location (i.e. the resonance wavelength) can be tuned over a wide spectral range, covering the entire source bandwidth, by changing the angle of incidence of the incoming beam with respect to the RPCS (Fig. 3a) [16].

Representative experimental results are shown in Figures 3-6. Figure 3a presents various notch-shaped excitation pulse spectra, obtained by changing the angle of the RPCS with respect to the excitation beam. The notch has a measured spectral width of 1.3nm FWHM (20cm$^{-1}$), and a rejection of >17dB. Figure 3b shows raw measured CARS spectra from toluene at two slightly shifted notch spectral locations. In both measurements, multiple blue-shifted notch features corresponding to toluene vibrational levels appear in the CARS spectrum. Two such features are apparent in each of the traces of Figure 3b, corresponding to the 787cm$^{-1}$ and 1005cm$^{-1}$ modes of toluene, respectively.

The origin of these blue-shifted vibrationally-resonant features in the CARS spectrum can be analyzed using the expression for the nonlinear polarization producing the vibrationally resonant CARS signal. For a singly resonant Raman transition, the vibrationally resonant nonlinear polarization spectrum, $P_r(\omega)$, driven by the pulse electric field, $E(\omega)$, can be approximated by [5,7]:

$$P_r^{(3)}(\omega) \propto G \int_0^\infty d\Omega \frac{E(\omega-\Omega)}{(\Omega_{vib}-\Omega)+i\Gamma} A(\Omega) \qquad (1)$$

where $\hbar\Omega_{vib}$ is the vibrational level energy, having a bandwidth $\Gamma$, and Raman strength G. $A(\Omega) = \int_0^\infty d\omega E^*(\omega-\Omega)E(\omega)$ is the second-order polarization driving the molecular vibration, i.e. the vibrational excitation amplitude at the frequency $\Omega$.

Since the notch filter is spectrally narrow compared with the pulse bandwidth, it has a negligible effect on the total pulse energy and temporal shape, and consequently on the vibrational excitation spectrum $A(\Omega)$. However, the notch shaping has a significant effect on the resonant CARS spectrum given by $P_r(\omega)$ (Eq.1). Due to the resonant term in the denominator in Eq.1, the main contribution to $P_r(\omega)$ originates from $E(\omega-\Omega_{vib})$. As a result, the resonant CARS spectrum has a shape that is similar to the excitation spectrum but blue-shifted by the vibrational resonant frequency. Therefore, the vibrationally resonant CARS spectrum contains narrow spectral dips at locations dictated by the resonant vibrational levels and the notch location.

The *measured* CARS signal is a coherent sum of the vibrationally resonant CARS spectrum given in Eq.1, and a broad featureless nonresonant polarization, $P_{nr}(\omega)$. When using femtosecond pulses for excitation, $P_{nr}(\omega)$ is typically orders of magnitude larger than the resonant field, and therefore the measured spectral intensity can be approximated by [5, 10]:

$$I_{meas}(\omega) \propto \left|P_{nr}(\omega) + P_r(\omega)\right|^2 \approx \left|P_{nr}(\omega)\right|^2 + 2\left|P_{nr}(\omega)\right|\left|P_r(\omega)\right|\cos(\phi(\omega)) \qquad (2)$$

Since the strong nonresonant signal is a spectrally smooth varying function, it is used to enhance the resonant signal by serving as a local oscillator in a homodyne detection scheme, given by the mixing term in Eq. 2 [5,10,13]. Due to the relative spectral phase between the resonant and nonresonant signals, $\phi(\omega)$, which originates from the vibrational resonance lineshape and the phase structure of the RPCS notch, the interference dips in the measured CARS spectrum appear at the longer wavelength end of each vibrational feature, and an interference peak manifest on its blue-side (Fig. 3b).

Although previous works have demonstrated the ability of resolving the vibrational spectrum from a single-shot measurement of the CARS spectrum [19-22], we find that the rapid tunability of the RPCS notch wavelength can be used to implement a differential measurement scheme, eliminating spectral artifacts caused by the spectral transfer function of the system. In a single-shot acquisition, a signal proportional to the resonant signal, $P_r(\omega)$, can be obtained from the measured CARS spectrum, by subtracting and normalizing by the nonresonant background (Eq.2):

$$P_r(\omega) \propto \frac{I_{meas}(\omega) - I_{nr}(\omega)}{P_{nr}(\omega)} \qquad (3)$$

where the nonresonant signal, $I_{nr}(\omega)=|P_{nr}(\omega)|^2$, which dominates the CARS spectrum is approximated by fitting a smooth curve to the measured spectrum, allowing for extraction of the resonant term [19,20]. Utilizing the RPCS notch tunability we effectively compensate for irregularities in the CARS spectrum arising from filter spectral transmission, spectrometer response, or non-smooth excitation spectrum, by taking the difference between two spectral measurements corresponding to slightly shifted notch locations. After further normalization by the vibrational excitation amplitude $A(\omega)$, we obtain a signal clean from the experimental system spectral artifacts, with peaks corresponding to the Raman lines strength, with accurate relative intensity values (Fig 3c):

$$I_{resolved}(\omega) = \frac{1}{A(\omega-\omega_{pr})} \left[ \frac{I_1^{meas}(\omega)}{P_{nr}(\omega)} - \frac{I_2^{meas}(\omega)}{P_{nr}(\omega)} \right] \quad (4)$$

where $I_1(\omega)$ and $I_2(\omega)$ are two raw measured CARS spectra corresponding to the slightly shifted probe locations, and $\omega_{pr}$ is the probe frequency (Fig3b,c).

Experimentally resolved vibrational spectra from various liquid phased samples and mixtures are shown in Figures 3 and 4. The vibrational spectra resolved via Eq.4 are in good agreement with the corresponding Raman spectra found in the literature (gray bars). We also verified that the raw measured CARS spectra are in good agreement with numerical simulations of the total CARS signal (Fig.3b), by substituting in equations 1 and 2, the measured excitation pulse spectrum and the Raman lines locations and relative intensity magnitudes found in literature.

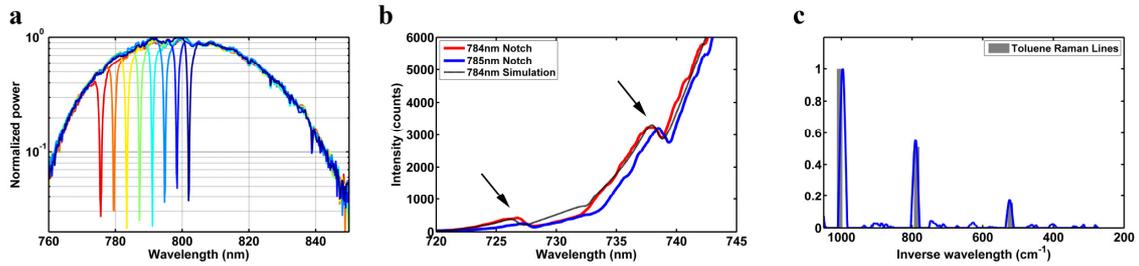

Fig. 3: Experimental results: (a) Several RPCS notch-shaped excitation spectra. The notch location is continuously tuned by the RPCS angle relative to the excitation beam; (b) Measured CARS spectra from toluene at two slightly shifted notch locations. In each measurement, sharp peak-and-dip interference features, corresponding to the vibrational resonances, appear in the CARS spectrum (marked by arrows), blue-shifted from the notch location by the vibrational frequency. (c) Resolved vibrational spectrum of toluene retrieved from (b) using Eq. 4 (peak normalized), with the known Raman lines depicted in gray.

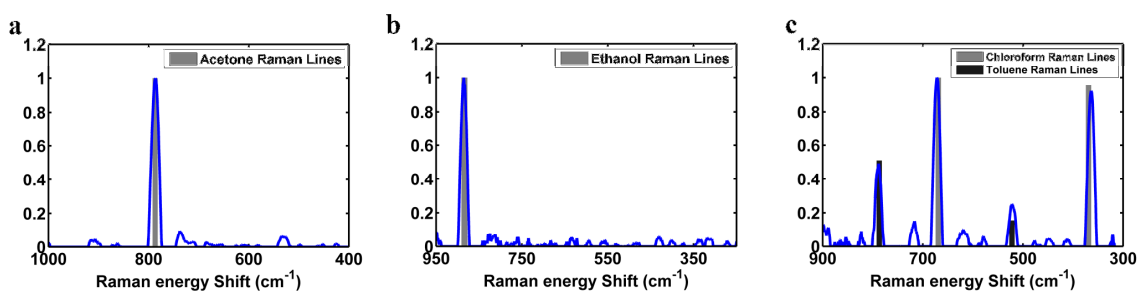

Fig. 4: Resolved vibrational spectra from pure samples of acetone (a), ethanol (b), and a 25% chloroform / 75% toluene mixture (c), obtained using the RPCS single-pulse CARS technique. The known Raman lines of the samples are depicted in gray. Spectra shown are peak-normalized.

RPCS notch-shaped single-pulse CARS can be used for microscopic vibrational imaging of unlabeled samples. Depth resolved micro-spectroscopic images of compound samples are achieved by scanning the sample using a piezo-controlled translation stage. Two examples are presented in Figures 5 and 6. Figure 5 shows the results from a thin sample comprised of a mixture of water and perfluorodecalin (Sigma-Aldrich P9900). The vibrational spectrum at each spatial location in the sample is resolved, and chemically specific contrast from a band of characteristic vibrational lines is obtained (Fig.5b). The resolved spectra allow easy discrimination and identification the two materials (Fig.5c). A similar study performed on a thin slice of a white potato is presented in Figure 6. The starch granules within the potato cells can be easily discriminated, by the characteristic 474cm$^{-1}$ skeletal mode of starch appearing their vibrational spectrum.

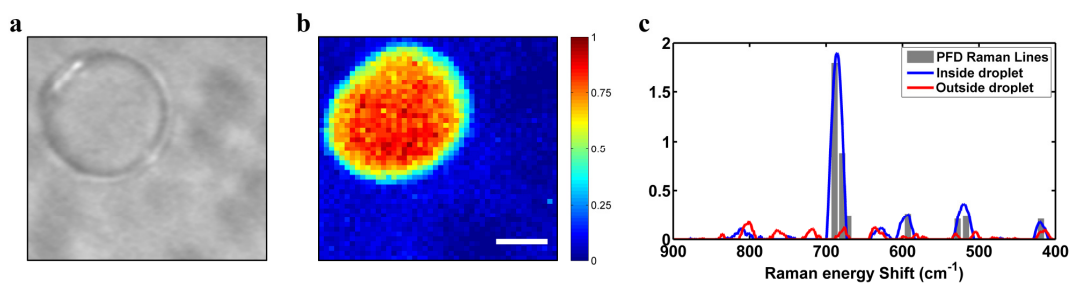

Fig. 5: Single-beam vibrational micro-spectroscopy of a mixture of water and perfluorodecalin using RPCS shaped single-pulse CARS: (a) transmission image; (b) Vibrational contrast image based on the 685cm$^{-1}$ band of perfluorodecalin (Scale bar 10μm); (c) Spatially resolved vibrational spectra reveal the vibrational spectrum of perfluorodecalin inside the droplet.

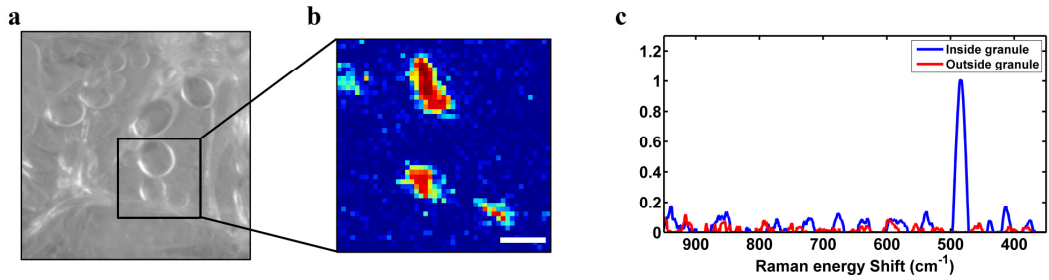

Figure 6: Single-beam vibrational micro-spectroscopy of potato cell starch granules using RPCS shaped single-pulse CARS: (a) potato slice transmission image; (b) corresponding vibrational contrast image based on the characteristic 474cm$^{-1}$ skeletal mode of starch (Scale bar 10μm); (c) Spatially resolved vibrational spectra inside and outside a granule, reveals the starch spectrum which is confined within the granules.

**Discussion**

In this study, we present an easily implementable and cost-effective shaper-less scheme for single-pulse single-beam CARS spectroscopy and microscopy. We demonstrate rapid RPCS notch-shaped pulse CARS micro-spectroscopy covering the Raman spectral range of 300-1000cm$^{-1}$ with spectroscopic resolution on the order of 20cm$^{-1}$, both of which can be easily improved. The lower and upper spectral limits are dictated by the detection filters and the pulse bandwidth, respectively. A dramatic increase in the spectral range can be achieved by an increased excitation bandwidth, as was recently shown using an octave-spanning femtosecond pulse source allowing for an upper limit of >4500cm$^{-1}$ [11]. The lower spectral limit can be extended down to the 100cm$^{-1}$ range by using sharper and carefully tuned excitation and detection filters [23]. The demonstrated spectroscopic resolution of 20cm$^{-1}$ is dictated by the chosen RPCS notch bandwidth, which is of the order of the vibrational linewidths of liquid-phase samples. Nonetheless, higher spectroscopic resolution is possible as the RPCS notch bandwidth can be designed to be as narrow as 0.1nm (~1.5cm$^{-1}$) [16], albeit reducing the probe pulse energy.

Since the RPCS is a monolithic and robust pulse shaping element, the implementation of single-pulse CARS using an RPCS results in a simple, stable, and virtually alignment free experimental system, employing a single beam and a single source of light. It is worth noting that several other single-source CARS techniques have been recently demonstrated, employing spectral focusing of two chirped femtosecond pulses [24-26], or using unchirped double-pulse excitation [20,27,28]. All of these schemes however, require a multi-beam and/or an interferometrically stable experimental apparatus, to achieve similar resolution and spectral coverage as the RPCS single-beam technique. A novel single-pulse technique recently introduced by Milner *et al*. [29] enables the retrieval of spectroscopic information by autocorrelation of a CARS spectrum generated by a noise-shaped excitation pulse. This method, however, recovers only the molecular beat frequencies, and requires averaging over several different spectra measurements for obtaining a signal to noise comparable to the RPCS notch-shaped multiplex CARS technique presented in this work.

An additional advantage of using an RPCS as a pulse shaping element is the ability for rapid pulse shaping. Modulation at kilohertz rates are obtained by using a galvanometric mirror mounted RPCS, while megahertz rates have been demonstrated by electro-optical modulation of the RPCS waveguide refractive index [16]. These high rates surpass the current refresh rate of the common liquid-crystal spatial light modulators (SLMs) by one to four orders of magnitudes, and hold a great potential for microscopy and remote-sensing applications, reducing the sensitivity to fluctuations in the probed sample or excitation pulses. In this work, we utilize the rapid tunability of the RPCS to remove spectral artifacts, using a differential measurement scheme. The resolved spectral line-shapes obtained with this method are similar to the line-shapes acquired in other differential-CARS techniques [24], and are not identical to the Raman line-shapes. Other, more involved spectral analysis techniques may be used to retrieve the true Raman line shapes from the measured CARS spectrum [10,20-22,30].

It is worth noting that a host of shaping techniques could be employed to generate a narrow spectral feature in the excitation spectrum, and obtain a CARS spectrum in a similar fashion to the RPCS notch-shaping. Examples for such techniques include notch interference filters, Fabri-Perot or Gires-Tournois interferometers, fiber Bragg gratings, and even a thin wire in the spectral plane of a prism/grating pulse compressor, a common apparatus in amplified femtosecond sources. Although these techniques do not possess all of the above mentioned advantages and simplicity of the RPCS shaping technique, the resultant CARS spectrum will be of similar nature, and will follow the analysis of equations 1-4.

In summary, we have demonstrated an easily implementable scheme for single-pulse, single-beam coherent vibrational spectroscopy and microscopy. The entire vibrational spectrum is resolved in a multiplexed single-shot measurement, using a simple experimental setup that contains a single light source and three optical elements which require straightforward alignment. The RPCS notch-shaping scheme can be directly extended from the infrared spectral range to the UV region, where classical liquid crystal light modulators are limited by absorption. The shaper-less femtosecond CARS scheme is compatible with other nonlinear techniques such as second- and third-harmonic generation and multiphoton fluorescence, and is a promising technique for practical spectroscopic applications.

**Materials and methods**

The RPCS is a commercial h124RE BioChip from Unaxis Balzers, Liechtenstein (optics.unaxis.com). The h124RE consists of a glass substrate (1.1mm thick Schott AF45, refractive index, n =1.52 at 800nm) with a uniformly etched sub-micron diffraction grating, having a period of 360nm and depth of 40nm. A subsequent thin (150nm) film layer of $Ta_2O_5$ (n=2.09 at 800nm) and a second identical etched grating. We illuminated the RPCS at classical incidence (i.e. the plane of incidence is perpendicular to the grating grooves) using TM polarization ('p' polarization, electric field within the plane of incidence). The advantages of TM polarization are the narrower resonant spectral bandwidth (notch width) in comparison to the TE resonance, and a larger effective spectral tuning range (300-900nm) as the background transmittance of the sample is higher for large angles due to the Brewster angle (57 degrees for our RPCS), thus improving the contrast of the notch. The laser source is a home-built prism-compensated Ti:Sapphire oscillator with a bandwidth of 60nm FWHM (940$cm^{-1}$), central wavelength of 800nm, and 200mW average power at 80MHz repetition rate. All filters used are Omega optical (AELP779, AGSP770, 3RD720-760). For vibrational imaging, samples were point scanned at 1μm steps, with 150ms pixel dwell time. Average laser power was ~30mW in all measurements except for the potato results of Figure 6, where the power was lowered to ~10mW. Figure 3 spectra were obtained using 120ms integration time. Figure 4a-c spectra were obtained with 500ms to 2s integration time, Figure 5c spectra with 1s integration time, and Figure 6c with 4s integration time.

**Acknowledgements**
We thank Silvia Soria (CNR, Italy) for generously providing the RPCS filters used in this work, and the AFM measurement of Figure 2c [17]. This work was supported by grants from the Israel Science Foundation and the Israel Ministry of Science.